\documentclass{aa}
\usepackage{graphicx}
\graphicspath{{./figures/}}
\usepackage[varg]{txfonts}
\usepackage{natbib}
\bibpunct{(}{)}{;}{a}{}{,}
\usepackage{hyperref}
\usepackage[svgnames]{xcolor}

\hypersetup{colorlinks=true, citecolor=Teal}

\newcommand*{\eg}{e.g.\@\xspace}
\newcommand*{\ie}{i.e.\@\xspace}
\newcommand*{\cf}{cf.\@\xspace}

\newcommand{\tausco}{$\tau$~Sco\@\xspace}
\newcommand{\etacar}{$\eta$~Car\@\xspace}

\newcommand{\mesa}{\mbox{\textsc{Mesa}}\xspace}
\newcommand{\arepo}{\mbox{\textsc{Arepo}}\xspace}

\defcitealias{schneider2019a}{Paper~I}

\newcommand{\comment}[1]{}

\begin{document}

\title{Magnetic field generation in mergers of massive main-sequence stars}

\author{Sebastian~T.~Ohlmann\inst{1}\fnmsep\thanks{\email{sebastian.ohlmann@mpcdf.mpg.de}} \and
       Fabian~R.~N.~Schneider\inst{2,3} \and
       Friedrich~K.~R\"{o}pke\inst{3,4,2} \and
       R\"{u}diger~Pakmor\inst{5} \and 
       Philipp~Podsiadlowski\inst{6} \and
       Volker~Springel\inst{5}
       }

\titlerunning{Magnetic field generation in mergers of massive main-sequence stars}
\authorrunning{S.~T.~Ohlmann et al.}

\institute{%
    Max Planck Computing and Data Facility,
    Gie\ss{}enbachstra\ss{}e 2, 85748 Garching, Germany
  \and
    Heidelberger Institut f\"{u}r Theoretische Studien,
    Schloss-Wolfsbrunnenweg 35, 69118 Heidelberg, Germany
  \and
    Zentrum f\"ur Astronomie der Universit\"at Heidelberg,
    Astronomisches Recheninstitut, M\"{o}nchhofstr. 12-14, 69120
    Heidelberg, Germany
  \and
    Zentrum f\"ur Astronomie der Universit\"at Heidelberg,
    Institut f\"ur Theoretische Astrophysik, Philosophenweg 12,
    69120 Heidelberg, Germany
  \and
    Max Planck Institute for Astrophysics,
    Karl-Schwarzschild-Straße 1, 85748 Garching, Germany
  \and
    University of Oxford, St Edmund Hall, 
    Oxford, OX1 4AR, United Kingdom
}

\date{Received 21/01/2026; accepted 15/06/2026}

\abstract{%
Magnetic fields are found in many astrophysical objects, ranging from galaxy clusters to the interstellar medium of galaxies and neutron stars. Strong surface magnetic fields are also observed in about 7\% of OBA-type stars, and stellar mergers are the likely origin of at least some of them. 
We investigated magnetic-field amplification during the merger of a $9$ and an $8\,M_\odot$ main-sequence star using 3D magnetohydrodynamic simulations from our previous work. We focused on the magnetic-field amplification mechanisms, field geometry, and the structure and properties of the resulting merger, in particular its rotational configuration. 
The merger produces a star–torus structure in which the core of the initially less massive star is surrounded by material from the primary. The torus contains ${\approx}\,3\,M_\odot$ and stores about 60\% of the initial orbital angular momentum, and dynamic mass loss is negligible. 
Initially, turbulent motions driven by Kelvin–Helmholtz and magneto-rotational instabilities generate small-scale (${\approx}0.1\,R_\odot$) magnetic fields. Subsequently, large-scale ordered azimuthal flows drive a larger-scale dynamo that amplifies and redistributes the magnetic energy to larger spatial scales (${\approx}5\,R_\odot$), producing a remnant threaded by a strong large-scale magnetic field. The final magnetic configuration consists of intertwined poloidal and toroidal components, with a residual small-scale structure that resembles previously identified stable magnetic field equilibria. The amplification process is largely insensitive to the initial binary separation, numerical resolution, and seed magnetic-field strength, indicating that the final field loses memory of the initial magnetic conditions.
The central regions of the merger remnant rapidly approach solid-body rotation, transitioning to a Keplerian-like profile within the surrounding torus. Our results support stellar mergers as a viable pathway for the formation of strongly magnetic massive stars and potentially highly magnetized compact remnants, such as magnetic white dwarfs and magnetars.
}


\maketitle

\section{Introduction}
\label{sec:introduction}

The merger of two stars is a common phenomenon in stellar astrophysics and can
involve compact objects such as neutron stars and black holes, but also stars
during their main nuclear-burning cycles. Whereas the former are observed, for
example, in gravitational-wave events \citep{abbott2016a,abbott2017b}, the
latter lead to outbursts called luminous red novae; V1309 Sco and V838 Mon are
two such prototypical events \citep{munari2002a, tylenda2011a}. Merged stars can
be rejuvenated \citep[\eg,][]{hellings1983a, braun1995a, dray2007a,
schneider2016a} and can appear as blue stragglers in star clusters and stellar
associations \citep[\eg,][]{hills1976a}. Some of the most massive stars known
might in fact be merger remnants \citep{schneider2014a}, and the Great Eruption
of \etacar in the 1840s that created the Homunculus Nebula is suggested to
be the result of a merger event \citep[\eg,][]{smith2018a, hirai2021a}. Some
merged stars may also give rise to unusual supernova progenitors such as the
surprising blue supergiant that led to supernova 1987A
\citep{podsiadlowski1989a, podsiadlowski1990a, podsiadlowski1992a, morris2007a,
menon2017a}, while others are thought to be the progenitors of interacting and
even superluminous supernovae \citep{justham2014a, schneider2025b}. Still, much
is unknown about the merging phase itself and the evolution and final fates of
merged stars. \citet{schneider2025a} present a general introduction to stellar
mergers and the related common-envelope phase, and \citet{schneider2025c}
reviews stellar mergers in depth.

Mergers can be induced in various ways \citep[see][]{schneider2025c}: by
dynamical encounters in dense stellar systems such as star clusters (\cf
``head-on collisions''), by the orbital decay of binaries in dense interstellar
or circumbinary media such as in disks of active galactic nuclei or during the formation of stars, by orbital dynamics in triple-star systems and other multiples (\eg, the von Zeipel--Lidov--Kozai mechanism), and by mass transfer in isolated binary stars. The mass transfer channel is supposed to lead to mergers for about 25\% of all massive stars \citep[$\gtrsim 10\,M_\odot$;][]{sana2012a} meaning that -- when integrating over a constant star-formation rate in a stellar population like in the Milky Way -- about 10\% of all massive stars are expected to be remnants of mergers from isolated binary evolution \citep{podsiadlowski1992a, demink2014a}. The binary evolution computations of \citet{henneco2024a} show that at least 16\% of massive, mass-transferring binaries will merge, and this fraction goes up to ${>}\,38\%$ if fully conservative mass transfer is assumed \citep{henneco2025a}.

Ever since the discovery of surface magnetic fields in the A-type star 78~Vir by \citet{babcock1947a}, the origin of strong surface magnetism in massive OBA stars has been debated. It has long been speculated that strong and large-scale magnetic fields are produced in stellar mergers and that they can persist over the long evolutionary timescales of stars \citep[see, \eg,][]{ferrario2009a, wickramasinghe2014a}. Today, direct simulations of essentially all types of stellar mergers and other dynamic stellar phases show the efficient amplification of magnetic fields. This is the case in mergers of binary white dwarfs \citep[\eg,][]{zhu2015a, pakmor2024a}, white dwarfs and neutron stars \citep[\eg,][]{moran-fraile2024a}, binary neutron stars \citep[\eg,][]{kiuchi2024a}, low- and high-mass main-sequence (MS) stars (\eg, \citealt{schneider2019a}, hereafter \citetalias{schneider2019a}; \citealt{ryu2024a,Vynatheya2025}), and common-envelope events \citep[\eg,][]{ohlmann2016b, ondratschek2022a, vetter2024a, gagnier2024a}. These strongly amplified fields are typically produced by small-scale instabilities, and it remains uncertain how and if the observed long-lived and large-scale fields emerge in these dynamic phases.

Mergers of MS stars offer a way to explain the origin of the strong surface
magnetic fields observationally found in about 7\% of OBA-type stars
\citep{landstreet1992a, donati2009a, fossati2015a, scholler2017a, grunhut2017a}.
For example, a merger origin can explain the magnetic stars \tausco
(\citealt{schneider2016a}; \citetalias{schneider2019a}), HR~2949
\citep{schneider2016a}, and HD~148937 \citep{frost2024a}. All three show clear
rejuvenation that can be understood within a merger scenario, and HD~148937 is
additionally surrounded by a bipolar nebula that appears to be the smoking gun
for debris ejected during a merger event. In the case of \tausco, we conducted
3D magnetohydrodynamic (MHD) simulations of the merger of MS stars with initial
masses of $9\,M_\odot$ and $8\,M_\odot$, and we show that magnetic
fields are strongly amplified and that the post-merger evolution of the star can explain many
of the observed properties of \tausco \citepalias{schneider2019a}. If the merger-generated
magnetic fields can persist until the pre-supernova stage of \tausco, the iron
core-collapse of the progenitor likely leads to the formation of a strongly
magnetic neutron star, a so-called magnetar \citep{varma2023a}.

Here, we extend and complement the analysis of the magnetic field amplification process in the $9+8\,M_\odot$ MS star merger presented in \citetalias{schneider2019a} and focus on the magnetic field structure, the emergence of large-scale magnetic fields, and the spin evolution of the merger remnant. In particular, we show how small-scale magnetic fields generated by the magneto-rotational and Kelvin--Helmholtz instabilities are converted into large-scale magnetic fields via large-scale fluid motions in the rotating post-merger star. We describe our computations in Sect.~\ref{sec:methods} and present our main findings in Sect.~\ref{sec:results}. The results are discussed in Sect.~\ref{sec:discussion}, and we conclude in Sect.~\ref{sec:conclusions}.

\section{Numerical methods}
\label{sec:methods}

The original simulation was presented in \citetalias{schneider2019a}, and we briefly summarize the applied computational set-up and methods here. The initial conditions for the merger
simulations in this paper are from stellar models computed with the stellar evolution
code \mesa (Sect.~\ref{sec:methodsevolution}) and represent a merger induced by isolated binary star evolution. The hydrodynamic simulations of
the merger are carried out using the moving-mesh MHD code \arepo
(Sect.~\ref{sec:methodshydro}).

\subsection{Stellar evolution calculations}
\label{sec:methodsevolution}

An $8\,M_\odot$ and a $9\,M_\odot$ MS star are
evolved to an age of 9\,Myr without rotation, using the stellar evolution code
\mesa \citep[version 9793;][]{paxton2011a,paxton2013a,paxton2015a}. The
computations used an initial helium mass
fraction of $Y = 0.2703$ and solar-like metallicity $Z\,{=}\,0.0142$
\citep{asplund2009a}. We used the Ledoux criterion for convection and
exponential convective core overshooting with a parameter of
$f_\text{ov}=0.019$. Mixing-length theory \citep{henyey1965a} was used with a
mixing-length parameter of $\alpha_\mathrm{MLT}=1.8$ and a semi-convective
efficiency of $\alpha_\mathrm{sm}=1.0$, but we did not apply the so-called MLT++
scheme in \mesa. We used the so-called ``Dutch'' wind mass-loss prescription.

\subsection{Magnetohydrodynamic simulations}
\label{sec:methodshydro}

We carried out the hydrodynamic simulations using the finite-volume
moving-mesh code \arepo \citep{springel2010a} with updates that improve the
convergence \citep{pakmor2016a}\footnote{The \arepo public version is available
at \url{https://gitlab.mpcdf.mpg.de/vrs/arepo} \citep[see][]{weinberger2020a}.}.
\arepo solves the ideal MHD equations using an HLLD
Riemann solver \citep{miyoshi2005a} and the Powell scheme for handling the
divergence constraint \citep{powell1999a}; for more details and tests of the
implementation, see \citet{pakmor2011d} and \citet{pakmor2013b}. We coupled the
OPAL equation of state \citep{rogers1996a,rogers2002a} to the hydrodynamics,
which is the same as used in \mesa in the relevant part of the
temperature--density plane.

In a first step, stable single star models were generated following the relaxation method as
described in \citet{ohlmann2017a} and \citetalias{schneider2019a}. Their stability
is analyzed in Appendix~\ref{sec:initialmodels}.
We give some parameters of the two initial stellar models in
Table~\ref{tab:stellarmodels}.

\begin{table}
  \centering
  \caption{Initial stellar models.}
  \label{tab:stellarmodels}
  \begin{tabular}{lccccc}
    \hline\hline
    Name & Mass & Radius & $t_\mathrm{s}\,^\mathit{a}$ & Luminosity & $T_\mathrm{eff}$ \\
     & ($M_\odot$) & ($R_\odot$) & (d) & ($10^3 L_\odot$) & $10^4$ K \\
     \hline
     MS-9M & 9.0 & 4.2 & 0.13 & 4.7 & 2.34 \\
     MS-8M & 8.0 & 3.8 & 0.12 & 3.1 & 2.21 \\
     \hline
     \multicolumn{6}{l}{\small $^\mathit{a}$ Sound crossing time}
  \end{tabular}
\end{table}

In a second step, the initial conditions for a merger were derived from these
stable stellar models. Setting up the system at Roche lobe overflow would
be computationally infeasible, so we added an
angular momentum loss term as an additional acceleration given as
$\vec{a}_i = - \vec{v}_i/\tau$
for each cell $i$ with $\tau=1.5 \times 10^6$\,s. We ran the simulation
for about 1.5 orbits with the loss term and continued without it from there on.
For comparison, we started one model without the loss term after only half this
time, and it showed a very similar evolution.

The simulations use 3.4
to 4 million cells in the high resolution runs and about 0.43 to 0.45 million
cells in the low resolution runs. We set up the initial magnetic field as
a dipole field, similar to that in \citet{ohlmann2016b}, with a surface field of
1\,$\mu$G. We ran one additional simulation without magnetic fields for
comparison. We provide an overview of the models in Table~\ref{tab:simulations}.

\begin{table}
  \centering
  \caption{Overview of the simulation runs.}
  \label{tab:simulations}
  \begin{tabular}{cccc}
    \hline\hline
    Model & Resolution$^\mathit{a}$ & Initial separation & $\vec{B}$ configuration \\
     & ($10^6$) & ($R_\odot$) & \\
    \hline
    1 & 3.4--4.0 & 6.4 & dipole \\
    2 & 0.43     & 6.4 & dipole \\
    3 & 0.45     & 7.0 & dipole \\
    4 & 0.43     & 6.4 & --- \\
    \hline
    \multicolumn{4}{l}{\small $^\mathit{a}$ Mean number of cells}
  \end{tabular}
\end{table}

To track the origin
of material during the merger, we added two passively advected scalars to the
initial conditions. We set each of the
scalars to 1 inside one of the stars and 0 everywhere else. Because we advected the
scalars in the simulation, their final value gives the mass fraction of a cell
originating from the corresponding star.

\section{Results}
\label{sec:results}

A set of simulations was run to study how the magnetic field is amplified in
MS star mergers, and they were first presented in \citetalias{schneider2019a}. We briefly recap a few important aspects of the merger simulations and then expand the analysis. During
the merger, the simulations show that the more massive primary star is disrupted
around the secondary, forming a core--disk structure
(Sect.~\ref{sec:mergerprocess}). Magnetic fields are amplified in the
simulations, first on small scales and later reaching large-scale structures
(Sect.~\ref{sec:magneticfields}). These fields are mostly toroidal in the final
merger product, which shows a core--disk structure where the core originates
mainly from the secondary and the disk mainly from the primary star
(Sect.~\ref{sec:structureproduct}). During the evolution, the simulations show
almost no ejection of mass (Sect.~\ref{sec:ejecta}).

\subsection{Evolution of the merger}
\label{sec:mergerprocess}

\begin{figure}[tb]
  \centering
  \includegraphics{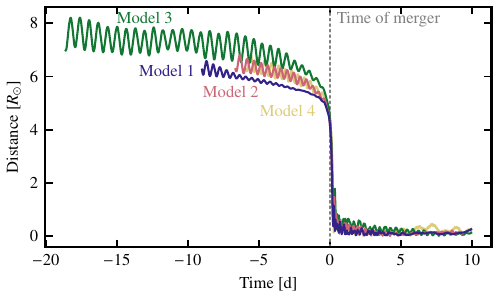}
  \caption{Evolution of orbital separation. The distance is computed between the
  centers of mass of the two stars. The center of mass of each star is computed by
  using the corresponding passive scalar as a weight.
  For the time of merger, we defined $t=0$ as the point in time at which the orbital
shrinkage becomes significant, \ie, where $a/\dot{a} < 1\,\mathrm{d}$ ($a$: semimajor axis).}
  \label{fig:distances}
\end{figure}

The simulations of the merger start after the orbit of the binary
system was shrunk to $6.4\,R_\odot$ (Model~1--3) or
$7.0\,R_\odot$ (Model~4), corresponding to orbital periods of about $0.46\,$d
and $0.52\,$d, respectively.
Strong mass transfer (${\approx}\,17\,M_\odot\,\mathrm{yr}^{-1}$) develops from the
more massive primary star onto the secondary. The orbital separation decreases
until the stars merge after roughly 23 (Model~1) to 35 (Model~4)
orbits (Fig.~\ref{fig:distances}). As expected, Model~4 with a larger
initial separation takes longer to merge than Models 1--3. Here, we defined $t=0$ as the point in time when $a/\dot{a} < 1\,\mathrm{d}$ (where $a$ is the semimajor axis).
Figure~\ref{fig:distances} also shows that the orbit is slightly eccentric, which is probably caused by the
system not being perfectly Keplerian after artificially
shrinking the binary orbit. The eccentricity is smaller for the high-resolution Model~1.

The evolution of the merger is described in \citetalias{schneider2019a} (see their Fig.~1).
In short, the primary star transfers mass through the Lagrange point L1 to the secondary star; mass and angular
momentum loss through L2 and L3 shrink the orbit, leading to the merger. The primary is disrupted around
the secondary with their cores merging at a later stage, leading to a symmetric merger product.
A core--disk structure forms with a core of about $3M_\odot$, dominated
by material from the secondary; the fraction of mass from the primary increases outward.

\subsection{Magnetic field amplification}
\label{sec:magneticfields}

\begin{figure*}[tb]
  \centering
  \includegraphics{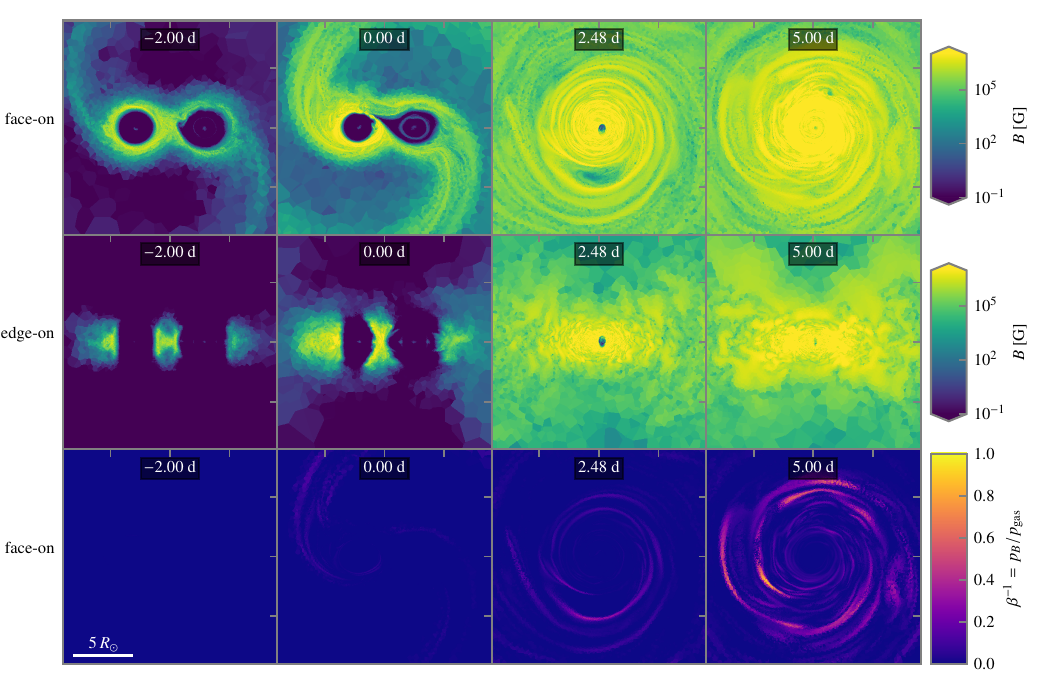}
  \caption{Evolution of magnetic fields for Model~1. Shown is the absolute value
  of the magnetic field in the orbital plane (\textit{top}) and perpendicular to
  the orbital plane (\textit{middle}). In the \textit{bottom} panel, the ratio of magnetic
  pressure to gas pressure indicates where the magnetic fields become
  dynamically relevant.
  For all pre-merger times ($t<0$), the plane is rotated such that the more
  massive star is located to the right.}
  \label{fig:magneticevolution}
\end{figure*}

During the rapid mass transfer leading up to the merger, magnetic fields are
strongly amplified. Due to the complex flow patterns, it is difficult to unambiguously
identify the amplification mechanisms. The maximum magnetic field strength grows exponentially for
some time over several orders of magnitude until it saturates at about
$10^8$\,G, independent of the different initial conditions
(see Fig.~2 in the extended data of \citetalias{schneider2019a}).
Fits to an exponential function yield amplification times of 0.2\,d to 1\,d.
The time evolution of the magnetic field in Fig.~\ref{fig:magneticevolution}
(first row)
shows that the initial amplification takes place in the region where the
accretion stream hits the secondary. Strong shear flows induce Kelvin-Helmholtz
instabilities, winding up the magnetic fields. The thickness of this layer is about
$0.8\,R_\odot$. In this region, we find that the flow
is unstable to the magneto-rotational instability (MRI; \citealp{balbus1998a}).
We estimated the length and timescale of the
fastest growing MRI channel in
our simulations \citep[\cf][]{rembiasz2016a}: this yielded a timescale of about 
0.5\,d and 
a wavelength of approximately $0.1\,R_\odot$, which is similar to the scales in our simulation
(timescale: 0.2--1\,d, length scale: $0.8 \,R_\odot$). Moreover, the strong amplification
of the magnetic field starts as soon as the wavelength of the fastest growing MRI channel is resolved on the grid.

When the stellar cores merge,
the magnetic field is distributed throughout the merger product, leading to a
nearly symmetric distribution of the field strength
(Fig.~\ref{fig:magneticevolution}, $t\,{>}\,0$\,d). Due to the differential rotation
during the merger process, the field structure is mostly toroidal: the fraction
of the toroidal magnetic energy is about $80-85\%$ of the total
magnetic energy (for more details, see Sect.~\ref{sec:structureproduct}).

The amplification slows and reaches saturation at later times, when
the ratio of magnetic to turbulent kinetic energy approaches the percent level
and comes closer to equipartition (see Fig.~3 in the extended data of \citetalias{schneider2019a}).
As a proxy for the turbulent kinetic energy, we used the kinetic energy in
the radial and $z$-directions, $E_{\mathrm{kin},r,z}$, which is not affected by
the large-scale angular motion. We even find super-equipartition, 
because some of the differential rotational energy from the large-scale ordered 
motion at $t\,{>}\,0\,\mathrm{d}$ is converted into magnetic energy.

\begin{figure}[tb]
  \centering
  \includegraphics{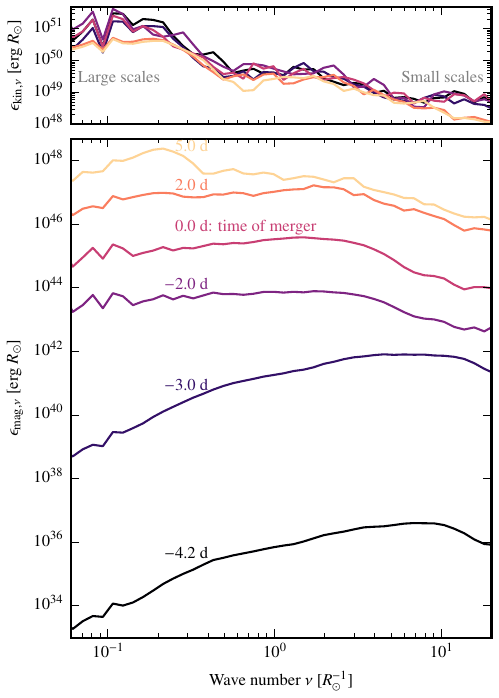}
  \caption{Power spectra of kinetic energy (\textit{top}) and magnetic energy
  (\textit{bottom}) at different times for Model~1. Shown is the energy per wave number
  as a function of wave number $\nu=k / 2\pi$ in inverse solar radii.
}
  \label{fig:powerspectra}
\end{figure}

The length scales of the magnetic field amplification can also be identified in power spectra of the kinetic and
magnetic energy (Fig.~\ref{fig:powerspectra}). Most of the kinetic energy
resides at large scales and low wave numbers for all times, as expected for the
rotation of the stellar cores and of the post-merger product
(Fig.~\ref{fig:powerspectra}, top panel). The power spectrum follows a
power law to smaller scales (exponent between $-1.4$ and $-1.1$), probably due to
turbulence seeded by flow instabilities. The exponents are smaller in magnitude
than the Kolmogorov exponent of $-5/3$ that would be expected for a homogeneous,
isotropic turbulent cascade. Plots of power spectra of the magnetic
energy (Fig.~\ref{fig:powerspectra}, bottom panel), however, show that magnetic
energy is injected at small scales (about $0.1$--$0.2\,R_\odot$) and propagates to
larger scales during the amplification (see $t\,{=}-4.2$\,d to $t\,{=}-2$\,d). After the
cores have merged, redistribution of the magnetic energy leads to a similar
spectrum at 5\,d as for the kinetic energy, with most of the energy at large
scales (about $3$--$10\,R_\odot$). The exponents of the power law were
obtained by a fit to $\epsilon_\mathrm{mag}$ for $\nu$ between 0.1 and
$10\,R_\odot^{-1}$ and are 1.38, $-0.20$, and $-0.78$ for $t=-4.2\,$d, $-2.0\,$d,
and $5.0\,$d, respectively. This behavior is similar to the MRI simulations by
\citet{bhat2016a}, in which large-scale modes only grow late in the evolution
when the amplification nears saturation. As mentioned there, large-scale
magnetic fields seem to appear when the amplification process saturates (see
also \citealp{ebrahimi2009a}). We suggest that this energy conversion from 
small to large scales is partially due to the large-scale ordered fluid motions 
winding up magnetic fields into a dominant toroidal component and an effective, 
large-scale dynamo that also explains the magnetic amplification beyond 
equipartition with the turbulent energy, akin to the case of white dwarf mergers studied 
in \citet{pakmor2024a}.

The impact of the magnetic fields on the dynamics of the merger can be
characterized by the ratio of magnetic pressure to gas pressure
$\beta^{-1}=p_{B}/p_{\mathrm{gas}}$ (see Fig.~\ref{fig:magneticevolution}, bottom row).
During the mass transfer leading up to the
merger, $\beta^{-1}$ stays below 1\%; thus, the effect of the magnetic fields on
the dynamics of the merger is very small. Only after the merger, the ratio of 
the magnetic pressure rises in some parts up to values between 0.3 and 1,
indicating that the magnetic fields have a stronger impact on the dynamics
of the merger product, especially on the differential rotation.

\subsection{Structure of the merger product}
\label{sec:structureproduct}

\begin{figure}[tb]
  \centering
  \includegraphics{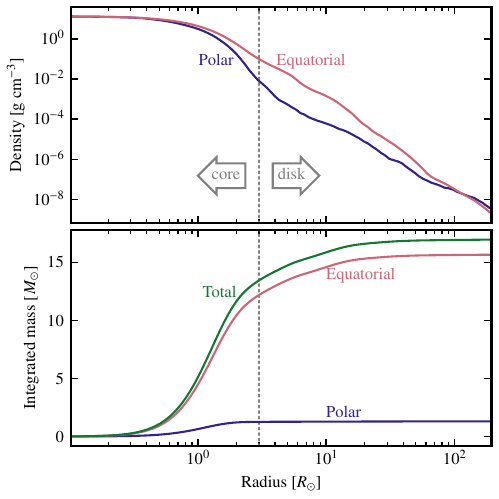}
  \caption{Characterization of the core and disk of the post-merger object at
    6\,d post-merger for Model~1. Shown are density (\textit{top}) and mass (\textit{bottom})
    over radius. Lines marked ``polar'' show cells with an angle to the orbital
    plane larger than $\pi/4$; those marked with ``equatorial'' show the other
    cells. The density was computed as the average over bins, and the mass was summed    over bins. The boundary between the core and the disk is located at approximately 
    $3\,R_\odot$ (dashed line).}
  \label{fig:diskcore}
\end{figure}

The large angular momentum in the binary system leads to the formation of a disk
around a rotating core in the merger product. The density profile
shows strong asymmetries between polar and equatorial regions, indicating that
the boundary between core and disk is located roughly at $3\,R_\odot$
(Fig.~\ref{fig:diskcore}). Moreover, the mass profile increases
only in the equatorial region for radii above $3\,R_\odot$
(Fig.~\ref{fig:diskcore}, bottom panel). Thus, almost all mass for radii above
$3\,R_\odot$ is contained in the equatorial region, i.e., in a disk-like
structure. At $t=4.44$\,d, the core contains a mass of about $13.8\,M_\odot$, and the
disk a mass of about $3.2\,M_\odot$. Moreover, the disk comprises 58\% of the
total angular momentum.

\begin{figure}[tb]
  \centering
  \includegraphics{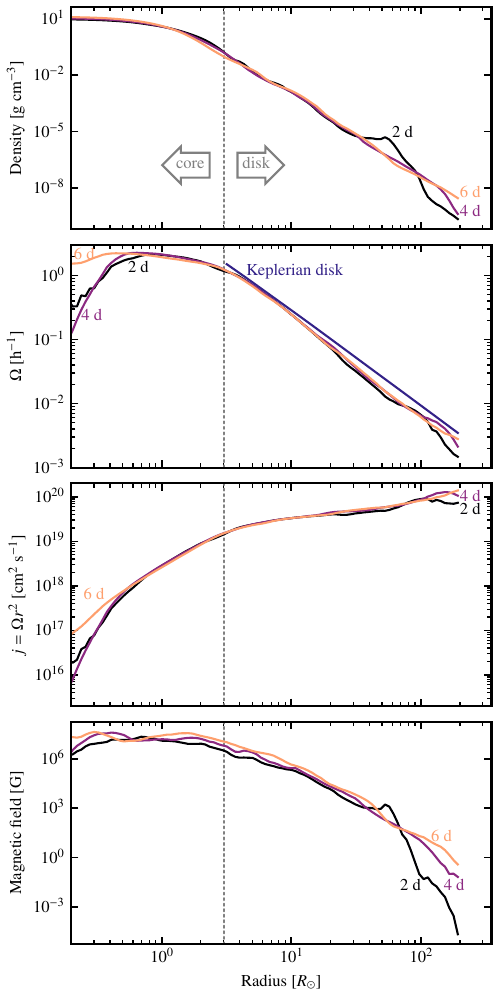}
  \caption{Density, angular velocity, specific angular momentum, and magnetic
    field profiles of the post-merger product at
  2\,d, 4\,d, and 6\,d. The core--disk boundary is indicated by the dashed line.
  For comparison, the angular velocity profile for a Keplerian disk is plotted.}
  \label{fig:profiles}
\end{figure}

Spherically averaged profiles of density, angular velocity, angular momentum,
and magnetic field
strength are shown in Fig.~\ref{fig:profiles}. 
The density profile remains very
similar between $t=4.44$\,d and $t=6$\,d (Fig.~\ref{fig:profiles}, top panel). At
early times after the merger ($t=2$ and $4.44$\,d), the angular velocity profile
reaches a maximum at some larger radius and is smaller in the core. Then,
angular momentum redistribution leads to a nearly solid-body rotation in the
central part ($\lesssim 2\,R_\odot$) at $t=6$\,d (Fig.~\ref{fig:profiles}, second row). In the outer part of the disk, however, no major angular momentum redistribution is visible (Fig.~\ref{fig:profiles}, third row)

The velocity at the core--disk boundary is about 700\,km\,s$^{-1}$ at
$t=4.44$\,d, \ie, 76\% of the critical Keplerian rotational velocity. As can be seen in
Fig.~\ref{fig:profiles} (second panel) the rotation curve in the disk always
stays sub-Keplerian. After six days, this disk has completed about 170 rotations at the base and about 0.3 rotations at the edge. The ratio of kinetic energy to gravitational energy
$T/|W|$ decreases from 0.07 at $t=4.44$\,d to 0.05 at $t=10$\,d and stays well below the
limit of 0.14 for non-axisymmetric instabilities in rotating fluids
\citep{ostriker1969a,chandrasekhar1970a,friedman1975a,lai1995a}.
Thus, no secular or dynamical, non-axisymmetric instabilities of the
configuration are expected.

The magnetic field strength (Fig.~\ref{fig:profiles}) still increases throughout
the merger product at the end of the simulation, which means that the amplification
is not yet completely saturated. Thus, there still seems to be an active slow dynamo,
probably connected to the large-scale rotation of the merger product. Ultimately,
the simulation would need to be continued to follow the amplification until
it is completely saturated. It is conceivable that the magnetic fields will grow so much that magnetically driven, bipolar outflows develop as are seen in simulations of similar binary systems \citep{ondratschek2022a, moran-fraile2024a, pakmor2024a, vetter2025a, Vynatheya2025}.

As already seen during the evolution of the merger (cf.\  \citetalias{schneider2019a}, Fig.~1),
the very central part of the merger product
originates from the secondary star, whereas the primary star is disrupted around
the core.
Only the inner $\sim2\,M_\odot$ of the product is
dominated by material from the secondary; the rest of the core stems from both
stars in similar proportion. The disk, however, is predominantly formed from material
of the primary star. This distribution is expected from the entropy and thus buoyancy profiles
of the progenitor stars: due to the similar mass and hence similar
evolutionary stage, the entropy is similar and mixing is expected over large
parts of the profile. Because the secondary has a lower mass, its
entropy is slightly smaller and thus its material dominates the central part.
This behavior is similar to the case of direct collisions as explored by
\citet{glebbeek2008a} and \citet{glebbeek2013a}, especially their case S.

\begin{figure*}[tbp]
  \sidecaption
  \centering
  \includegraphics{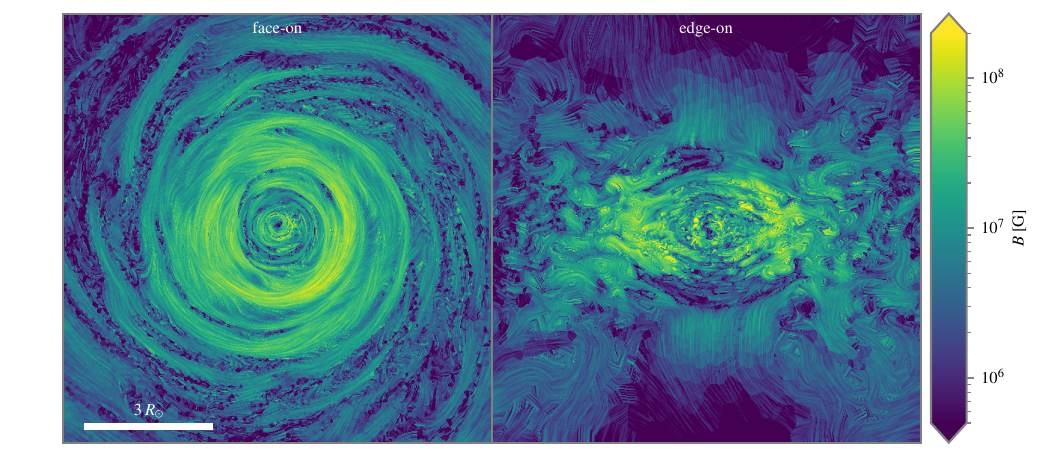}
  \caption{Magnetic field structure of the merger product at $t=6$\,d shown
    using line integral convolution. The colors indicate the magnitude of the magnetic
    field. The field lines are visualized in each plane
    by computing a convolution of a white noise texture and a kernel along the
    field lines.
    Figure from \citet{schneider2025c}.}
  \label{fig:magneticstructure}
\end{figure*}

The magnetic field structure of the final merger product is visualized in
Fig.~\ref{fig:magneticstructure}. The field structure is mostly toroidal; the
energy of the toroidal magnetic field component is about 80--85\% of the total
magnetic energy. In this range, magnetic field configurations are expected to be
stable in stars \citep{braithwaite2004a, braithwate2006a}. Moreover, the configuration remains stable
after the final merger, and this time already corresponds to a few
Alfvén timescales. However, in the regime of fast rotation ($\Omega \gg \omega_\mathrm{A}$), the evolutionary timescale of the magnetic field is $\tau_\mathrm{evol}\sim \tau_\mathrm{A}^2 \Omega$, where $\tau_\mathrm{A}$ is the Alfvén timescale \citep{braithwaite2013a}. Hence, significant changes in the magnetic field are not expected to be observed over the timescales simulated here after the field has reached an equilibrium configuration.

\begin{figure*}[tb]
  \centering
  \includegraphics{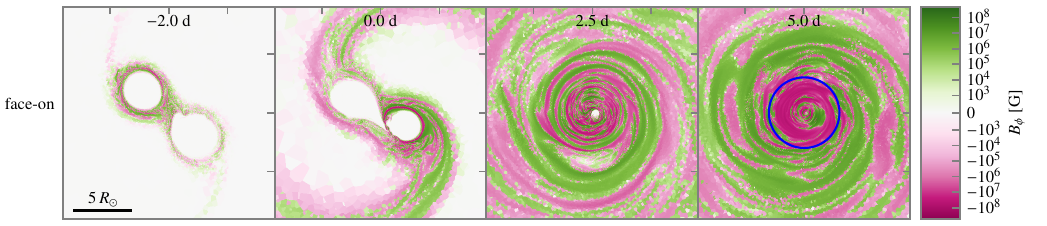}
  \caption{Evolution of the polar component of the magnetic field 
  for Model~1. Shown is $B_\phi$ in the orbital plane. The color scale is symmetric and logarithmic around
  0; the inner part between $-10^{3}$ and $10^{3}$ is linear.}
  \label{fig:bphi}
\end{figure*}

Due to Ohmic resistivity, the magnetic field is expected to dissipate and
the corresponding timescale depends on the square of the coherence scale $R$ of the
magnetic field. Using Spitzer resistivity, we can estimate the dissipation
timescale as
\begin{equation}
    \tau \approx 20\ \mathrm{Gyr} \left(\frac{R}{R_\odot}\right)^2 \left(\frac{T}{10^6\ \mathrm{K}}\right)^{\frac{3}{2}},
    \label{eq:spitzer}
\end{equation}
where $T$ is the temperature of the plasma. In Fig.~\ref{fig:bphi}, we plot the
azimuthal component of the magnetic field, which shows that the magnetic field is
coherent on length scales of roughly $1 R_\odot$. With temperatures between 
$10^5\ \mathrm{K}$ and $10^7\ \mathrm{K}$, we get timescales between $0.7\ \mathrm{Gyr}$ and $700\ \mathrm{Gyr}$, larger than the lifetime of the star, which
should rather be about $10\ \mathrm{Myr}$. Hence, the magnetic field of the
star should
remain present at least until it reaches the MS. Modeling the
subsequent stellar evolution in more detail
shows that the merger product could evolve into an MS star with strong
magnetic fields, similar to \tausco (\citetalias{schneider2019a}; \citealt{schneider2020a}).

The merger process leads to a denser and hotter core than a comparable
16.9\,M$_\odot$ MS star in full equilibrium: heat is mostly generated in the
central region, leading to a larger internal energy in the core.
The angular momentum redistribution
leads to high rotational velocities outside the core
(Fig.~\ref{fig:profiles}), and thus the kinetic energy is higher than the
internal energy outside the core, especially in the thick disk.
This is related to the fact that the
angular momentum of a binary increases outward ($J\propto \sqrt{a}$), whereas
the orbital energy decreases outward ($E_\mathrm{orb}\propto a^{-1}$).
This redistribution of energies is important to predict 1D merger products that
can be used for further stellar evolution modeling (see also
Sect.~\ref{sec:comparison}).

Similarly, the rotational configuration of the merged star is crucial for further evolution. Right after the merger (\eg, $t=2\,\mathrm{d}$, the toroidal structure rotates at sub-Keplerian angular velocities $\Omega$, and $\Omega$ reaches a maximum at radii of $0.5\text{--}1.0\,R_\odot$ (Fig.~\ref{fig:profiles}). Hence, these inner regions are stable against the MRI. With continuing evolution, the innermost regions spin up through angular momentum transport into these regions, and the entire central merger remnant approaches solid-body rotation that transitions into the sub-Keplerian angular velocity of the disk-like structure at the core--disk transition radius of about $3\,R_\odot$. Such an angular-velocity rotational profile is the minimum-energy configuration and appears to be reached within a few dynamical timescales after the merger.

\subsection{Dynamical ejecta}
\label{sec:ejecta}

Because the further evolution of the merger product and its optical display
might be influenced by dynamical ejecta, we computed how much mass is unbound at
the end of the simulation. To this end, we summed the mass of all cells in the
simulation for which the total energy is positive, i.e., for which the sum of
potential, kinetic, and internal energies is positive. This is an upper bound
because it assumes that the internal energy can be converted completely to kinetic energy
and thus neglects other effects, such as cooling.
After the merger settled into the core--disk structure, at $t=6$\,d,
about $0.02\,M_\odot$ of material is unbound, corresponding to 0.14\% of the
total mass of the system. Not much mass is ejected compared to other stellar
mergers (e.g., common envelope phases; see \citealp{ohlmann2016a}) because
the binding energy of the system is still quite high, about
$-4\times 10^{50}\,\text{erg}$. The unbound mass moves outward with an average
radial velocity of about 180\,km/s and is predominantly ejected near the orbital
plane with a distribution that shows some tails to the polar directions.

\section{Discussion}
\label{sec:discussion}

\subsection{Impact of numerical parameters}
\label{sec:parameters}

The main numerical parameters that affect the simulation outcome are the
resolution, the initial distance and the magnetic field configuration.
As already discussed in \citetalias{schneider2019a}, we assessed the impact of resolution
by running Model~1 with about 10 times more
cells than Model~2 (see Table~\ref{tab:simulations}), which roughly corresponds to a factor of 2 in each
spatial dimension. 
Despite differences in the evolution of the magnetic field strength, the final field strength 
and configuration are quite similar, which
demonstrates that they are robust with respect to the resolution.

The initial distance influences the outcome as more angular momentum is present
in the system for larger initial distances. In principle, the simulation should
start at Roche-lobe overflow; however, this is prohibitively expensive in
computational time because the mass transfer progresses only slowly in the
beginning, leading to a long time to the actual merger. As already described, we
removed angular momentum artificially to obtain a starting configuration
that leads to a merger more quickly. In reality, the system would have more
angular momentum, leading probably to a longer mass loss via L2 and L3 in the
phase before the actual merger. The largest impact of this is probably on the
detailed structure of the outer part of the disk. To assess the impact of a
larger initial separation, we ran Model~3 with a larger initial distance. As one
can see in Fig.~\ref{fig:distances}, it takes longer until the cores finally
merge, but the overall evolution is similar. Also, the magnetic field evolution
(Fig.~\ref{fig:magneticevolution}) shows a slightly different shape, but the
final field strength is of similar magnitude and configuration. Thus, we expect that the
main conclusions are not affected by the detailed choice of this parameter, but
that it mainly affects the outermost structure of the disk.

In addition, we ran a simulation without magnetic fields at all to assess their
impact on the orbital evolution (Model~4). The orbital evolution is very similar
to Model~2 with the same resolution (see Fig.~\ref{fig:distances}); thus, the
impact of the magnetic fields on the dynamical evolution is small. Moreover, the ratio of
magnetic pressure to gas pressure is only about 1\% in the phase leading up to
the merger. This indicates that the magnetic fields do not greatly influence 
the dynamics of the gas in this phase.

\begin{figure*}[tb]
  \centering
  \includegraphics{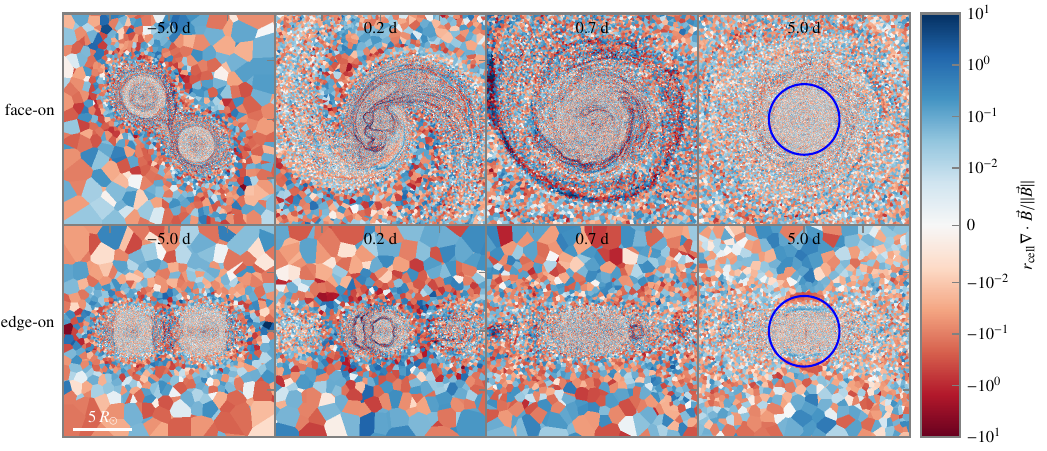}
  \caption{Evolution of the relative error in the divergence of the magnetic
    field ($r_\mathrm{cell} \nabla\cdot \vec{B} / \|\vec{B}\|$) for Model~1
    in the orbital plane (\textit{upper row}) and perpendicular to it (\textit{lower row}).
    The color scale is symmetric and logarithmic around 0; the inner part
    between $-10^{-2}$ and $10^{-2}$ is linear.
  }
  \label{fig:divergence}
\end{figure*}

A typical numerical issue in MHD simulations is fulfilling the
divergence constraint $\nabla\cdot\vec{B}=0$. In principle, the continuous equations of MHD
preserve this constraint for initial conditions fulfilling it. However,
discretization errors also lead to errors in $\nabla\cdot\vec{B}$. As explained
in \citet{pakmor2013b}, we used the Powell method \citep{powell1999a} to
control the errors in the divergence. We verified that the errors in our
simulations behave very similarly to those in the tests shown there (see
Fig.~\ref{fig:divergence}): averages over
larger regions are small and typically adjacent cells show different signs in
the error, indicating that the large-scale impact of these local errors is
small. The volume-weighted average of the relative divergence error is
$7.0\times10^{-4}$, $-1.6\times10^{-4}$, $8.9\times10^{-3}$, and
$6.2\times10^{-2}$ for the times given in Fig.~\ref{fig:divergence}.

\subsection{Comparison to other simulations}
\label{sec:comparison}

\citet{glebbeek2013a} have conducted a series of stellar merger simulations as head-on
collisions and subsequent stellar evolution calculations for a range of masses
and evolutionary points. They used a smoothed-particle hydrodynamics code with $130\,000$ particles per star,
whereas we used a grid-based approach with 1 to 2 million cells per star and included MHD.
In their simulations, the equal-mass collisions eject
between 4\% and 8\% of the total mass (see their Fig. 3), whereas in our
simulation, only 0.14\% of the mass is unbound (compare with Sect.~\ref{sec:ejecta}).
This might be due to the difference in the initial conditions as material can
escape more easily in head-on collisions, whereas in our merger simulations a
large disk forms that is bound.  In our simulation, the core is dominated by
material from the secondary star,
corresponding to ``case S'' in \citet{glebbeek2013a}. As explained there, this
can be attributed to the lower buoyancy in the secondary star. They find ``case
S'' for relatively unevolved primaries at moderate mass ratios, which would also
apply to our case. 

\citet{ryu2024a} simulated parabolic head-on collisions of low-mass stars with
nonzero impact parameters similar to those mentioned above, but they also used
the moving-mesh MHD code \arepo and found strong magnetic field amplification.
Such collisions contain less angular momentum than mergers in the circular
binaries studied here and thus form not-so-clear massive disk-like structures in
the merger remnant. The magnetic field amplification mechanism is otherwise in
qualitative agreement with that found in our mergers of circular high-mass
binaries: the head-on collisions invoke several contact or grazing encounters before the final coalescence, and, in each encounter, small-scale dynamos amplify the magnetic fields. From the work presented in \citet{ryu2024a}, however, it remains unclear whether large-scale magnetic fields are formed. \citet{Vynatheya2025} compare head-on collisions of massive stars with binary mergers of the same stars. They find systematic differences between the two scenarios; in particular, they find that merger remnants have signs of an ordering of the magnetic field similar to our simulation, which is absent for collision remnants. They focus on a qualitative analysis of the final state of their collision and merger remnants. The latter seems consistent with our simulation. For further differences between binary mergers and head-on collisions, see, for example, \citet{schneider2025c} and \citet{Vynatheya2025}.

Remarkably, the magnetic field amplification process in mergers of MS stars is
very similar to that in mergers of white dwarfs and neutron stars presented in
\citet{pakmor2024a} and \citet{kiuchi2024a}, respectively. In these binary
white-dwarf and neutron-star mergers, small-scale magnetic fields are also
produced by a turbulent dynamo, and magnetic energy is then transferred to
larger scales by a large-scale, MRI-driven dynamo. A similar situation is also
found in neutron star--white dwarf mergers \citep{moran-fraile2024a} and
common-envelope phases of low-mass and high-mass stars \citep{ondratschek2022a,
vetter2024a}.
Despite the similarities, all these simulations feature outflows that are magnetically driven, whereas our simulations do not show such outflows. As explained above, the magnetic field throughout the merger structure keeps increasing. Hence, such outflows might also show up in our merger model if the simulations are run sufficiently long.

\section{Conclusions}
\label{sec:conclusions}

We have studied in detail the magnetic-field amplification in the merger of a $9$ and an $8\,M_\odot$ MS star in an isolated binary system and the properties of the resulting merged star. The simulations were first described in \citetalias{schneider2019a} to explain observations of the magnetic massive star \tausco, and we extended the analysis in this work. The focus of this study was on the magnetic field amplification process, the magnetic field geometry, the mixing during the merger, and the rotational configuration of the resulting merger product. The main findings can be summarized as follows.
\begin{itemize}
    \item We see the formation of a star--torus structure in which the core of the initially less massive $8\,M_\odot$ secondary star is in the center and the initially more massive $9\,M_\odot$ primary star is distributed around it. The material in the torus originates mostly from the $9\,M_\odot$ primary. The rotationally supported torus has a mass of ${\approx}3\,M_\odot$ and holds ${\approx}\,60\%$ of the initial angular momentum of the binary.  
    \item There is hardly any dynamic mass loss, and only ${\approx}\,0.14\%$ of the total mass is ejected.
    \item We observe a robust magnetic-field amplification of more than ten orders of magnitude. The magnetic-field growth is initially sourced by turbulent energy in the merger and then by large-scale, ordered azimuthal fluid motions. In the end, the magnetic field energy reaches a super-equipartition level with respect to the kinetic energy in the radial and $z$-direction, \ie, our proxy of the turbulent energy. The outcome appears robust and does not depend on the initial separation of the binary, the numerical resolution, or the strength of the seed magnetic field. We thus conclude that the memory of the initial seed magnetic field is lost.
    \item The magnetic field is initially amplified by small-scale turbulent
      dynamos invoking the Kelvin-Helmholtz instabilities (shear) and MRIs. Later, the large-scale, ordered azimuthal fluid motions drive another dynamo process that pumps the magnetic energy from small (${\approx}\,0.1\,R_\odot$) to large (${\approx}\,5\,R_\odot$) scales. The magnetic energy power spectrum in the final merger structure follows that of the kinetic energy, and the merger remnant is threaded by a large-scale magnetic field. 
    \item The final magnetic-field geometry is characterized by a large-scale poloidal and toroidal structure with a still non-negligible small-scale component. The field structure resembles that found by \citet{braithwaite2004a} and \citet{braithwaite2006a} to be stable over thermal timescales and possibly up to nuclear timescales.
    \item The spherically symmetric and pressure-supported central parts quickly approach solid-body rotation and transition into a Keplerian-like rotation profile in the torus. 
\end{itemize}
This work contributes to the understanding of the complex physics in stellar
mergers, but several questions remain unanswered. The stability of the amplified
magnetic fields needs to be assessed, and the further evolution of the
star--torus structure will determine the exact properties of merged stars such
as their rotational velocities and surface chemistry
\citep[see][]{schneider2020a, schneider2025c}. Addressing these questions is
essential to further understanding the properties of magnetic massive stars and
their possible highly magnetic white-dwarf and neutron-star remnants (\ie, polars and magnetars).

\begin{acknowledgements}
We thank the anonymous reviewer for their constructive and insightful report that greatly helped improve this manuscript. 
FRNS, FKR, RP, and VS were supported by the Klaus Tschira Foundation.
This work has received funding from the European Research Council (ERC) under the European Union’s Horizon 2020 research and innovation programme (Grant agreement No.\ 945806) and is supported by the Deutsche Forschungsgemeinschaft (DFG, German Research Foundation) under Germany’s Excellence Strategy EXC 2181/1-390900948 (the Heidelberg STRUCTURES Excellence Cluster). The work of FKR  is supported by the Deutsche Forschungsgemeinschaft (DFG, German Research Foundation) -- RO 3676/7-1, project number 537700965, and by the European Union (ERC, ExCEED, project number 101096243). Views and opinions expressed are, however, those of the authors only and do not necessarily reflect those of the European Union or the European Research Council Executive Agency. Neither the European Union nor the granting authority can be held responsible for them. 
For data processing and plotting, NumPy and SciPy
\citep{oliphant2007a}, IPython \citep{perez2007a}, and Matplotlib
\citep{hunter2007a} were used.
\end{acknowledgements}

\bibliographystyle{aa}
\bibliography{astrofritz}

\begin{appendix}
\section{Initial models}
\label{sec:initialmodels}

\begin{figure}[h]
  \centering
  \includegraphics{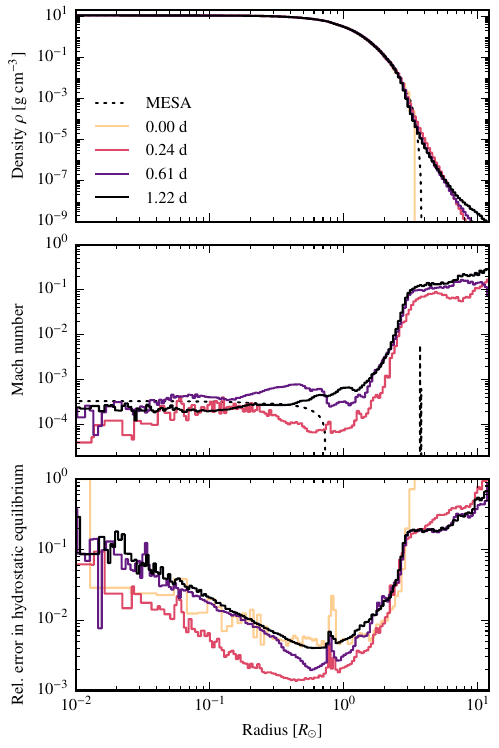}
  \caption{Initial model of the $8\,M_\odot$ MS star. From \textit{top
  to bottom}: Radially averaged profiles of the density, the Mach number, and the difference in the
  hydrostatic equilibrium (gradient of pressure and gravitational force) at
  different times during the relaxation run.}
  \label{fig:initialmodel}
\end{figure}

The initial MS star models
are stable in the hydrodynamic simulations for many dynamical timescales,
as can be seen in Fig.~\ref{fig:initialmodel}. The density profile
(top panel) remains the same in the interior of the
star; only the boundary smears out slightly, which is to be expected due to the
insufficient resolution in these low-density regions. For the same reason, the
Mach numbers are slightly larger at the edge of the star, while the Mach number in the
interior remains below $10^{-3}$ (middle panel).
Hydrostatic equilibrium ($\nabla p=\rho \vec{g}$) is fulfilled to
better than a few percent in most parts of the star
(bottom panel).

\end{appendix}

\end{document}